\documentclass[11pt,a4paper]{article}
\usepackage{amssymb}
\usepackage{amsthm,epsf}
\usepackage{amsfonts}
\usepackage{epsfig}
\usepackage{graphicx}

\newtheorem{remark}{Remark}

\begin{document}
\title {\bf
A Simple Model of Scale-free Networks  \\
Driven by Both Randomness and Adaptability% Linking and Anti-preferential \& Random Deletion
\footnote {This research is supported in part by the National
Natural Science Foundation of China through grant 70171059 and by
Hong Kong Research Grant Council through grant HKUST6189/01E}}
\date{}
\maketitle

\author{\centerline {\small Dinghua Shi$^{1}$
 , \ Xiang Zhu$^{2}$ \ and \ Liming Liu $^{2,}$\footnote [3] {The corresponding author: E-mail address: liulim@ust.hk}}}
 %Fax: (852) 2358 0062. E-mail address: Liulim@ust.hk}}}

\centerline{\small $^1$Department of Mathematics, College of
Science, Shanghai University,} \centerline{\small Shanghai 200436, China }
 \centerline{\small E-mail address: shidh2001@263.net}

\centerline{\small $^2$Department of Industrial Engineering and Engineering Management,}
 \centerline{\small Hong Kong University of Science and Technology,}
 \centerline{\small Clear Water Bay, Kowloon, Hong Kong}
 \centerline{\small E-mail address: zhuxiang@ust.hk}

% \centerline{\small E-mail address: liulim@ust.hk}

\abstract {In this paper, we present a simple model of scale-free
networks that incorporates both preferential \& random attachment
and anti-preferential \& random deletion at each time step. We
derive the degree distribution analytically and show that it
follows a power law with the degree exponent in the range of
$(2,\infty)$. We also find a way to derive an expression of the
clustering coefficient for growing networks and compute the
average path length through simulation.
\bigskip


\bigskip

}

\section{Introduction}
\ \ \ \   Complex networks play a crucial role in a wide range of
practical systems of technological, biological, and social importance
\cite{Albert02,Strogatz01}. For example, the
Internet, the World Wide Web (WWW), communities of scientists and
biological cells can all be described as complex networks.
Although various complex networks exist in various different fields,
their evolutions are driven by a few rules.
We believe that three intrinsic rules are behind the evolutions of most complex networks.
They are \emph{randomness}, \emph{adaptability} and \emph{hereditary}.
The existing investigations usually focus on one or two of the three rules.

The earliest study of complex networks can be traced to the
investigation of regular graphs characterized by a large
clustering coefficient and a long average path length. Erd\"{o}s
and R\'{e}nyi \cite{Erdos60} initiate the studies of complex
networks as random graphs. They propose the ER
model %in which given $N$ nodes, there exists a link between two
%nodes with a certain probability $p$. The ER model
which has a short average path length and a small clustering
coefficient. Later observations have found that some real networks
have not only short average path lengths like random graphs but
also large clustering coefficients like regular graphs. These two
features characterize the small-world network. Watts and Strogats
\cite{Watts98} later develop a model based on regular graphs in
which links are random rewired with a fixed probability. For some
range of small rewiring probabilities, their model successfully
displays the small-world characteristics.
%Namely, their model is called a small-world network, which represents an interesting development for the study of complex networks.
Two things are common for random graphs and small-world networks:
randomness of connections between nodes and exponential decay of
the tail of the degree distribution.

However, more recent empirical evidences from the Internet and
WWW, among other complex networks, show a fundamentally different
picture, i.e., the tail of the degree distribution follows a power
law. Two general features have been observed in many real-world
networks: successive additions of new nodes and preference to link
to the existing nodes. This shows that randomness is not the
unique feature of networks and leads to the introduction of
scale-free networks in 1999 by Albert, Baraba$\acute{a}$si, and
Jeong in their pioneering works
\cite{Albert99,Barabasi99a,Barabasi99b}, which start a new phase
in the study of complex networks. Albert, Baraba$\acute{a}$si, and
Jeong propose two mechanisms to characterize the evolution of a
scale-free network \cite{Barabasi99a,Barabasi99b}: \textbf{the
growth}, starting from $m_0$ nodes, the network grows at a
constant speed, i.e., adding one node at each time step and
connecting to $m ~ (m \leq m_0)$ existing nodes; \textbf{the
preferential attachment}, the chance that an existing node
receives a connection from a new node is proportional to the
number of connections it already has. Here the phenomenon of
preferential attachment reflects the \emph{adaptability} of
complex network. The authors show that, under these two
mechanisms, a network evolves into a stationary scale-free state.
Its degree distribution follows a power law with the degree
exponent $\gamma = 2.9\pm 0.1$ from simulation analysis and
$\gamma = 3$ from the analytical result. These results are
significant for complex networks and these two mechanisms become
the first model, referred to as the BA model. Although the BA
model can be used to interpret many phenomena of complex networks,
the degree exponent is a constant, which is a weakness since most
empirical studies shows that $\gamma$ can be either less than $3$
or large than $3$ in real complex networks \cite{Albert02}.

To improve the original BA model, many researchers suggest
different mechanisms for both growth and preferential attachment
under which the range of $\gamma$ varies from $2$ to infinity. In
the following, we will briefly review some significant works.
%For detail information, please refer to corresponding literature.

Krapivsky, Redner, Leyvraz \cite{Krapivsky00} examine the effect
of a nonlinear preferential connection probability $\Pi(k)$ on
complex networks. By analyzing the rate equation, they demonstrate
that the topology of the network is scale-free only when the
preferential attachment is asymptotically linear. Dorogovtsev,
Mendes, and Samukhin \cite{Dorogovtsev00a} use a master-equation
approach to study complex networks in which the probability
$\Pi(k)$ is proportional to the sum of a node's initial
attractiveness and the number of incoming edges. By applying
mean-field theory, Dorogovtsev and Mends \cite{Dorogovtsev00c}
consider both preferential attachment and random removal (with
equal probability) in the evolution of a network. Albert,
Baraba$\acute{a}$si \cite{Albert00b} study internal edges and
rewiring, Dorogovtsev and Mends \cite{Dorogovtsev00b} propose
models for gradual aging.
%The degree exponents for all these models lies in $(2,+\infty).$

Different from the BA model, Krapivsky, Rodgers, Render
\cite{Krapivsky01b} consider a growing network with directed
edges. In their model, at each time step, either a new node or a
new link is randomly added to the network and the attachment
probability depends on the in- or out-degrees of nodes. By solving
rate equations, they conclude that both in- and out-degree
exponents lies in $(2, \infty)$. Kleinberg et al.
\cite{Kleinberg99}, Kumar et al. \cite{Kumar00a,Kumar00b} address
an alternative preferential mechanism named copy mechanism by
adding random links with ``prototype'' nodes. It is found that the
copy mechanism is equivalent to a linear preferential attachment.
Krapivsky and Render \cite{Krapivsky01a}'s edge redirection
mechanisms is mathematically equivalent to Kumar et al.'s model
discussed above.

From the reviewed works, we find two common facts: (1) under
linear growth, the range of the degree exponent can be extended to
infinity by adding randomness into model; (2) local events and
growth constraints have a similar function, which is to make the
degree exponent vary between $2$ and $3$. Although the above
research extends the range of the degree exponent, their proposed
mechanisms are relatively complex. Compared with the above models,
Liu et. al \cite{Liu02}'s model is relatively simple. It combines
the ideas from \cite{Erdos60} and \cite{Barabasi99a} to model the
probability that a new node is connected to node $i$ already in
the network. They find that the degree exponent is no less than
$3$, so the model is not applicable to situations when the degree
exponent is between $2$ and $3$, which is the most common range
observed in real world. Recently, Chen and Shi \cite{Chen04}
introduces the concept of anti-preferential deletion into the BA
model and show that $2 < \gamma < 3$. This shows that integrating
randomness and anti-preferential deletion into the BA model, one
may construct a simple model for a general class of scale-free
networks with $\gamma>2$.

This research is mainly motivated by the above observation.
Based on \cite{Liu02} and \cite{Chen04}, we propose a simple evolution
model of complex networks with preferential and random attachment,
anti-preferential and random deletion. We show that the network
self-organizes into a scale-free network. We obtain \
the expression of the degree exponent analytically, and find it lies in $(2,\infty)$.
Clustering coefficient is another key network parameter,
but analytical estimations are hard to obtain for growth networks as reflected by the
current state of the literature. In this paper, we are able to
derive an analytical expression for the clustering coefficient.
Our method can be useful for similar studies.
In short, our model is constructed from simple mechanisms and can be applied
to analyze a general class of complex networks.

We organize the paper as follows. In the next section, we present
a simple model of scale-free network. In section 3, we obtain the
degree exponent analytically. Section 4 develops a method to derive the
clustering coefficient. Section 5 discusses the average path length.
We conclude the paper in Section 6 by pointing out some future
research opportunities.

\section{Model Description}
\ \ \ \ Our network starts with $m_0$ completely connected nodes.
At each time step, the following two procedures are performed:

(i) A new node is added to the system: $m(\leq m_0)$ new edges
from the new node are connected to $m$ different existing nodes. A
node $i$ with degree $k_i$ will receive a connection from the new
node with the linear-preferential probability
\begin{equation}
\label{eq:prep}
\Pi(k_{i}) = \frac{(1-p)k_{i}+p}{\sum_{j}[(1-p)k_{j}+p]}, %~~~  0 \leq p \leq 1
\end{equation}
where $p$ is the probability that the selection of an existing node (for attachment)
is random while $(1-p)$ is the probability that the selection of an existing node
(for attachment) preferential.
%(2) n new edges between old nodes are produced: a node i is selected as a end of a
%new edge, with the preferential probability    (ki).

(ii) $c$ old links are deleted: We first select node $i$ with at
least one link as one end of a link with the
anti-linear-preferential probability
\begin{equation}
\label{eq: antip} \Pi^*(k_i)=N^{-1}(t-1)[1-\Pi(k_i)],
\end{equation}
where $N(t-1)$ is the number of connected nodes with nonempty
links at $t$ time step and $N^{-1}(t-1)$ is used as the normalized
coefficient such that $\sum_i\Pi^*(k_i)=1.$ Then, we choose
another node $j$ from the neighborhood of node $i$ (denoted as
$O_i$ ) as the other end of the link with probability
$K_i^{-1}\Pi^*(k_j)$, where $K_i=\sum_{j\in O_i }\Pi^*(k_j)$. We
delete the link connecting nodes $i$ and $j$. We repeat this
procedure $c$ times to delete $c$ old links.

The basic ideas of the above process is to use a linear combination of
a random selection probability and a preferential selection probability
as the selection probability.
We believe that this linear selection rule for attachment and deletion models
real world networks more closely, from the point of view of the evolutionary theory.
%, where $m\geq 1$,$c\geq 0$, $m>c$.

\section{Degree Distribution}
\ \ \ \ By the continuum theory, $k_i(t)$ approximately satisfies
the following dynamic equation:
\begin{eqnarray}
\frac{\partial k_i}{\partial t}&=&m\Pi(k_i)-c\left[\Pi^*(k_i)+\sum_{j\in O_i}\Pi^*(k_j)K_j^{-1}\Pi^*(k_i)\right] \nonumber\\
&=& m \Pi(k_i)-c\left[\Pi^*(k_i)+\Pi^*(k_i)\sum_{j\in O_i}K_j^{-1}\Pi^*(k_j)\right] \nonumber\\
\label{eq: dk}
&\approx&m\frac{(1-p)k_{i}+p}{[2(1-p)(m-c)+p]t}-c\frac{2}{t},~~~\mbox{for
large $t$} \hspace{0.1cm}
\end{eqnarray}
where the last approximation comes from $\sum_j
[(1-p)k_j+p]=[2(1-p)(m-c)+p]t$ and $\{\Pi^*(k_i)+\Pi^*(k_i)\sum_{
j\in O_i}K_j^{-1}\Pi^*(k_j)\} \approx 2/t$. In the approximation,
we assume that $N(t-1) \approx t$ and $\sum_{j\in
O_i}K_j^{-1}\Pi^*(k_j) \approx 1$. Near the end of next session,
we give a simulation result to check the accuracy of degree
exponent obtained under this assumption.

Let $t_i$ be the time step when node $i$ is added to the network.
Initially, node $i$ has $k_i(t_i) = m$ links, thus the above
equation has the following solution:
\begin{equation}
\label{eq: so} k_i(t)=B \left [ \left ( \frac{t}{t_i} \right )^\beta -1 \right ] + m,\hspace{0.1cm}
\mbox{for large $t$}
\end{equation}
where the dynamic exponent
\begin{equation}
\label{eq: de}
\beta=\beta(m,p,c)=\frac{m(1-p)}{2(1-p)(m-c)+p},  %\hspace{0.1cm} \mbox{because $k_i(t)\propto t^{\beta}$,$\beta<1$}
\end{equation}
and the coefficient
\begin{equation}
\label{eq: co} B=B(m,p,c)=m+\frac{mp-2c[2(1-p)(m-c)+p]}{m(1-p)}.
\end{equation}

In the solution procedure, we require $0<\beta<1$ and $B>0$ for
the feasible solution. To guarantee $0<\beta<1$, the parameters should satisfy
\begin{equation}
\label{eq: conb}(1-p)(m-2c)+p>0.
\end{equation}

On the other hand, the condition $B>0$ holds if and only if
\begin{eqnarray}
m^2(1-p)+mp-4c(1-p)(m-c)-2cp&>&0 \Leftrightarrow\nonumber\\
(1-p)(m-2c)^2+p(m-2c)&>&0\Leftrightarrow\nonumber\\
(m-2c)[(1-p)(m-2c)+p]&>&0. \nonumber
\end{eqnarray}

Therefore, we conclude that $B>0$ if only if (\ref{eq: conb}) is
held. In sum, we see that $m>2c$ is a sufficient condition for
both $0<\beta<1$ and $B>0$.

Assume that $t_i$ follows the uniform distribution over interval
$(0, t)$. Then, by (\ref{eq: so}), we have
\begin{eqnarray}
P(k_i(t)<k)&=&1-(\frac{B}{B-m+k})^{1/\beta}\frac{t}{m_0+t},\nonumber \\
P(k,t)&=&\frac{\partial P(k_i(t)<k)}{\partial k}\nonumber \\
&=&\frac{t}{m_0+t}\frac{1}{\beta}B^{1/\beta}(k+B-m)^{-\gamma},\nonumber \\
P(k)&=&\frac{1}{\beta}B^{1/\beta}(k+B-m)^{-\gamma},~~~(t\rightarrow\infty),
\end{eqnarray}
where
\begin{equation}\label{eq: se}
\gamma=1+\frac{1}{\beta}=3+\frac{p-2c(1-p)}{m(1-p)}.
\end{equation}
Thus, this system self-organizes into a scale-free network with
a degree exponent given by (\ref{eq: se}).

Since $[p-2c(1-p)]/[m(1-p)]$ is increasing in $p$ and
$\gamma=3-2c/m>2$ for $p=0$, we have $\gamma>2$. In particular,
when $p\leq 2c(1-p)$, we can generate values of $\gamma$ between
$2$ and $3$. Such $\gamma$ values have been observed in different
networks including the WWW and movie actor collaboration networks
\cite{Barabasi99b}. For $p\rightarrow 1$, we have $\gamma
\rightarrow \infty$ while for $(1-p)(m-2c)+p \rightarrow 0$, we
obtain $\gamma \rightarrow 2$. Further, when $p=c=0$, it yields
the BA model \cite{Barabasi99a}; when $p=0$ and $c>0$, it gives
Model B (with $n=0$) proposed by Chen and Shi \cite{Chen04}; when
$p>0$ and $c=0$, it is equivalent to the model studied in Liu et. al \cite{Liu02}.

\begin{figure}[htbp]
\center{\psfig {file=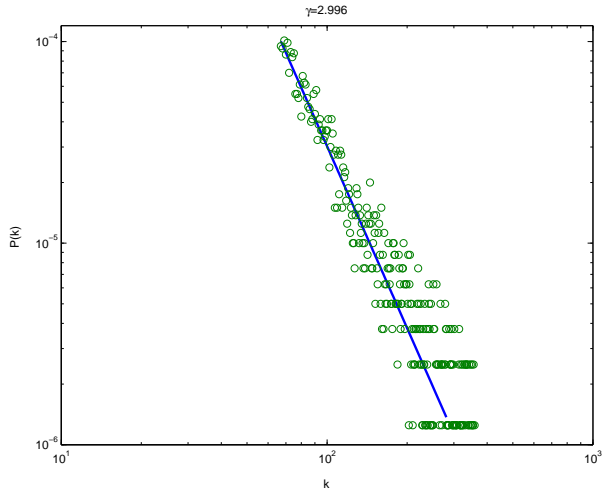, width=8cm}} \caption{the Degree
Distribution} \label{fig4}
\end{figure}

We now use simulation to compute the degree distribution of our
model. We set $m_0=10$, $m=5$, $c=1$ and $p=0.667$. Analytically,
$\gamma=3$ from (\ref{eq: se}). In the experiment, we take the
average from $100$ runs. After computation, we obtain
$\gamma\approx 2.996$, and the coefficient is $29.692$. Figure
\ref{fig4} shows that the results of the simulation, which
indicates the approximations in (\ref{eq: dk}) are reasonable.

\begin{remark}
\label{re: addlink} Suppose that at each time step, we also
perform an additional process: $n$ new edges between old nodes are
produced: a node $i$ is selected as a end of a new edge, with the
probability $\Pi(k_i)$ given by (\ref{eq:prep}). Then, the new
degree exponent is given by
\begin{equation}
\label{eq: add} \gamma=3+\frac{p-2(1-p)(n+c)}{(m+2n)(1-p)}.
\end{equation} We can show that this process has no impact on the
range of $\gamma$. Further, by (\ref{eq: add}), when we let $c=0$,
we find that the range of $\gamma$ is kept the same under the
effect of $n$, which indicates that the function of adding new
edges between old nodes is equivalent to that of
anti-linear-preferential deletion.
\end{remark}

\begin{remark}
Now, at each time step, we consider another additional operation:
we rewire $n$ existing edges in the network: select
randomly a node $i$ and a link $l_{ij}$ connected to it. Next we
rewire this link and replace it with a new link $l_{i^{\prime}j}$
that connects node $j$ and node $i^{\prime}$ chosen with the
probability $\Pi(k_i)$ given by (\ref{eq:prep}). This operation is
repeated $n$ times. As a result, the degree exponent is given by
\begin{equation}
\label{eq: rew} \gamma=3+\frac{p-2(1-p)(n+c)}{(m+n)(1-p)}.
\end{equation}
 Clearly, $n$ also does not affect the range of
$\gamma$. Moreover, if we let $c=0$, we see that the function of
rewiring old edges between old nodes is equivalent to that of
anti-linear-preferential deletion.
%for c=0, p>(n-m)(1-p) limit the range of n
\end{remark}

\section{Clustering Coefficient}
\ \ \ \ In this section, we present a method to derive an explicit
expression for clustering coefficients of growing network models.

Consider a node $l$. When the size of the network is $N$, there
will be $k_l(N)$ nodes in its neighborhood. The maximum possible
number of links among all the neighbors of node $l$ is
$k_l(N)[k_l(N)-1]/2$. The clustering coefficient $C_l(N)$ of node
$l$ is then defined as the ratio between the actual number of
links among all the nodes in the neighborhood and
$k_l(N)[k_l(N)-1]/2$. The clustering coefficient of the network is
then the average of the clustering coefficients of all the nodes
in the network.

To compute the clustering coefficient, we will rewrite dynamic
equation (\ref{eq: dk}) in the following integral form
\begin{equation}
\label{eq: in} k_i(t)= m+\int_i^t \frac{m(1-p)k_i(j)+mp-2ca_1}{a_1
j}dj,
\end{equation}
where $a_1=2(1-p)(m-c)+p>0$.

From (\ref{eq: in}), we find that the expected number of links
connecting node $i$ added at the $i$th time step with node $j$
added at the $j$th time epoch up to time step $t$ is given by
$$\int_i^t\frac{m(1-p)k_i(j)+mp-2ca_1}{a_1j}dj.$$

Next, by continuous theory, we obtain that the probability for the
existence of a link from the node $j$ to node $i$ ($i<j$), i.e.,
\begin{eqnarray}
Prob\{(ij)\}&=& \frac{m(1-p)k_i(j)+mp-2ca_1}{a_1j} \nonumber \\
\label{eq:probij}
&=& \frac{a_2 j^{\beta-1}}{a_1 i^{\beta}},
\end{eqnarray}
where the second equality is followed from (\ref{eq: so}) and $a_2=m^2(1-p)+mp-2ca_1>0$.

%where the approximation is reasonable since the product of $\Pi^*(k_{i})$ and $\Pi^*(k_{j}$

To find the number of actual connections among neighbors of a
given node $l$, we need to consider the sequence (age) by which
node $l$ and its neighbors appear. For example, $l<i<j$ means that
node $l$ is older than node $i$ which is in turn older than node $j$.
Then, the expected number of edges between node $i$ and
node $j$ that are neighbors of node $l$ is given by
$\int_l^Ndip(li)\int_i^N dj p(lj)p(ij).$ Similarly, we have to
count five other cases: $i<l<j$, $i<j<l$, $l<j<i$, $j<l<i$ and
$j<i<l$. The related integration expressions of six cases are
given in (\ref{eq:cep}), respectively. Note that we count the
links between any two node twice, we need to divide the sum of six
integrations by $2$.
Also, we approximate the maximum number of connections by $k_l(N)^2/2$.
Thus, we obtain
\begin{eqnarray}
 C_l(N)&=&\frac{1}{k_l(N)^2}\left[\int_l^N di
p(li)\int_i^N dj p(lj) p(ij)+\int_1^ldi p(il)\int_l^N dj p(lj) p(ij)\right.\nonumber\\
&&\left.+\int_1^ldi p(il)\int_i^l dj p(jl)p(ij)+\int_l^Ndi
p(li)\int_l^idjp(lj)p(ji)\right.\nonumber\\
\label{eq:cep} &&\left.+\int_l^N di p(li)\int_1^l dj
p(jl)p(ji)+\int_1^ldi p(il)\int_1^idj p(jl)p(ji)\right].
\end{eqnarray}

Now, we consider first two extreme cases of our model:
$\beta=0.5$ and $\beta=0$. For $\beta=0.5$, we set $p=0$, $c=0$.
In this case, our model is equivalent to the BA model, and $p(ij)=
p(ji)=m(ij)^{-1/2}/2$. Furthermore, (\ref{eq:cep}) can be
simplified to
\begin{eqnarray}
 C_l(N)&=&\frac{\int_1^Ndi p(li)\int_1^Ndj p(lj)p(ij)}{k_l(N)^2} \nonumber\\
&=&m(\ln N)^2/(8N)\nonumber\\
 \label{eq:ceo}
&\propto& \frac{(\ln N)^2}{N}.
\end{eqnarray}
The last equation is the same as the one provided in \cite{Klemm02}.
Noting that $C_l(N)$ is independent of $l$, (\ref{eq:ceo})
also gives the cluster coefficient of the whole network.
%When $\beta=1/2$, (\ref{eq: approc}) is rewritten as $$C(N)\propto \ln N /N.$$

When $c=0$ and $p=1$, $\beta =0$ and we have a random network.
We can rewrite (\ref{eq:probij}) as
\begin{eqnarray}
Prob\{(ij)\}
  \label{eq: probij1}
 &=&\frac{m}{j}.
\end{eqnarray}
The integrations of (\ref{eq:cep}) can be simplified,
\begin{eqnarray*}
\int_l^N di p(li)\int_i^N dj p(lj) p(ij)&=&\int_l^N di
p(li)\int_l^i
dj p(lj)p(ji)=m^3(l^{-1}-N^{-1}-\frac{\ln N-\ln l}{N}),\\
\int_1^l di p(il)\int_l^N dj p(lj) p(ij)&=&\int_l^N di
p(li)\int_1^l
dj p(jl)p(ji)=m^3\frac{l-1}{l}(l^{-1}-N^{-1}),\\
\int_1^l di p(il)\int_i^l dj p(jl)p(ij)&=&\int_1^l di
p(il)\int_1^i dj p(jl)p(ji)=m^3\frac{l-1-\ln
 l}{l^2}.
\end{eqnarray*}
Using $k_l(N)\approx m \ln N$ as in \cite{Liu02}, we can similarly obtain
\begin{eqnarray*}
 C_l(N)&=&m\left[l^{-1}-N^{-1}-\frac{\ln N-\ln l}{N}+\frac{l-1}{l}(l^{-1}-N^{-1})+\frac{l-1-\ln
 l}{l^2}\right]/(\ln N)^2,\\
&=&m\left[2(l^{-1}-N^{-1})-\frac{\ln N-\ln
l}{N}-\frac{l^{-1}-N^{-1}}{l}+\frac{l-1-\ln
 l}{l^2}\right]/(\ln N)^2.
\end{eqnarray*}
It is easy to see that
\begin{eqnarray}
\label{eq: beta} C(N)&=&\int_1^N C_l(N) d l /N
\propto\frac{1}{N\ln N}.
\end{eqnarray}
%Note that when $\beta=0$, (\ref{eq: approc}) is rewritten as $$C(N)\propto \ln N /N^2.$$
The analytical results obtained above for random networks are new.

For the general case, explicit formula for the clustering coefficient is more difficult to
obtain. The following analysis provides a good general approximation.

We need to compute the $6$ integrations in (\ref{eq:cep}) separately. For
$l<i<j$, we have
\begin{eqnarray*}
\int_l^N di p(li)\int_i^N dj p(lj) p(ij)&=&\int_l^N
\frac{a_2i^{\beta-1}}{a_1 l^{\beta}} di
\int_i^N\frac{a_2j^{\beta-1}}{a_1 l^{\beta}}\times \frac{a_2
j^{\beta-1} }{a_1 i^{\beta}} dj \nonumber\\
&=&\frac{a_2^3l^{-2\beta}}{a_1^3}\left[\frac{N^{2\beta-1}(\ln
N-\ln
l)}{2\beta-1}-\frac{N^{2\beta-1}-l^{2\beta-1}}{(2\beta-1)^2}\right].
\end{eqnarray*}

For $i<l<j$,
\begin{eqnarray*}
\int_1^l di p(il)\int_l^N dj p(lj)
p(ij)&=&\int_1^l\frac{a_2l^{\beta-1}}{a_1 i^\beta}di \int_l^N
\frac{a_2j^{\beta-1}}{a_1 l^\beta}\times
\frac{a_2j^{\beta-1} }{a_1 i^{\beta}} dj \\
&=&\frac{a_2^3l^{-1}}{a_1^3}\frac{(1-l^{-2\beta+1})N^{2\beta-1}-l^{2\beta-1}+1}{(2\beta-1)^2}.
\end{eqnarray*}

For $i<j<l$,
\begin{eqnarray*}
\int_1^l di p(il)\int_i^l dj
p(jl)p(ij)&=&\int_1^l\frac{a_2(l/i)^\beta}{a_1 l}di \int_i^l
\frac{a_2(l/j)^\beta}{a_1 l}\times\frac{a_2(j/i)^\beta}{a_1 j}dj\\
&=&\frac{a_2^3l^{2\beta-2}}{a_1^3}\left[\frac{\ln
l}{2\beta-1}+\frac{l^{-2\beta+1}-1}{(2\beta-1)^2}\right].
\end{eqnarray*}

For $l<j<i$,
\begin{eqnarray*}
\int_l^N di p(li)\int_l^i dj
p(lj)p(ji)&=&\int_l^N\frac{a_2i^{\beta-1}}{a_1 l^\beta}di \int_l^i
\frac{a_2j^{\beta-1}}{a_1 l^\beta}\times\frac{a_2i^{\beta-1}}{a_1
j^\beta}dj\\
&=&\frac{a_2^3l^{-2\beta}}{a_1^3}\left[\frac{N^{2\beta-1}(\ln
N-\ln
l)}{2\beta-1}-\frac{N^{2\beta-1}-l^{2\beta-1}}{(2\beta-1)^2}\right].
\end{eqnarray*}

For $j<l<i$,
\begin{eqnarray*}
\int_l^N di p(li)\int_1^l dj
p(jl)p(ji)&=&\int_l^N\frac{a_2i^{\beta-1}}{a_1 l^\beta}di \int_1^l
\frac{a_2l^{\beta-1}}{a_1
j^\beta}\times\frac{a_2i^{\beta-1}}{a_1 j^\beta}dj\\
&=&\frac{a_2^3l^{-1}}{a_1^3}\frac{(1-l^{-2\beta+1})N^{2\beta-1}-l^{2\beta-1}+1}{(2\beta-1)^2}.
\end{eqnarray*}

For $j<i<l$,
\begin{eqnarray*}
\int_1^l di p(il)\int_1^i dj
p(jl)p(ji)&=&\int_1^l\frac{a_2l^{\beta-1}}{a_1 i^\beta}di \int_1^i
\frac{a_2l^{\beta-1}}{a_1 j^\beta}\times\frac{a_2i^{\beta-1}}{a_1
j^\beta}dj\\
&=&\frac{a_2^3l^{2\beta-2}}{a_1^3}\left[\frac{\ln
l}{2\beta-1}+\frac{l^{-2\beta+1}-1}{(2\beta-1)^2}\right].
\end{eqnarray*}

%{\bf Dividing (?)} the summation of the above results by $2$,
Putting the summation together,
we obtain the expectation number of actual connections among
neighbors of a given node $l$:
\begin{eqnarray}
E&=&(\frac{a_2}{a_1})^3\left\{\frac{N^{2\beta-1}[l^{-2\beta}(2\beta-1)(\ln
N-\ln l)-2l^{-2\beta}+l^{-1}]}{(2\beta-1)^2}\right.\nonumber\\
\label{eq: sum} &&\left.+\frac{l^{2\beta-2}(2\beta-1)\ln
l+3l^{-1}-2l^{2\beta-2}}{(2\beta-1)^2}\right\}.
\end{eqnarray}
Substituting (\ref{eq: sum}) and $k_l(N)^2$, noting (\ref{eq:
so}), into (\ref{eq:cep}), we obtain
\begin{eqnarray}
 C_l(N)&=&2(\frac{a_2}{a_1})^3 \left\{\frac{N^{2\beta-1}[(2\beta-1)(\ln
 N-\ln
 l)-2+l^{2\beta-1}]}{(2\beta-1)^2[BN^{\beta}-(B-m)l^{\beta}]^2}\right.\nonumber
\\\label{eq:cea}
&&\left. +\frac{l^{4\beta-2}(2\beta-1)\ln
 l+3l^{2\beta-1}-2l^{4\beta-2}}{(2\beta-1)^2[BN^{\beta}-(B-m)l^{\beta}]^2}\right\}.
\end{eqnarray}

Analytical integration of the above equation is next to impossible in general.
For an approximation of the network clustering coefficient and to identify
it asymptotic behavior as $N$ becomes large, it is reasonable to
approximate $k_l(N)$ by $B(N/l)^\beta$. This allows us to rewrite (\ref{eq:cea}) as
\begin{eqnarray}
C_l(N)\label{eq:ceaa}& \approx&
2(\frac{a_2}{a_1})^3\frac{N^{2\beta-1}[(2\beta-1)(\ln
 N-\ln l)-2}{B^2(2\beta-1)^2N^{2\beta}} \nonumber \\
&&
 +\frac{l^{2\beta-1}]+l^{4\beta-2}(2\beta-1)\ln
 l+3l^{2\beta-1}-2l^{4\beta-2}}{B^2(2\beta-1)^2N^{2\beta}}.
\end{eqnarray}
Taking average on both sides of (\ref{eq:ceaa}), it is easy to conclude that
\begin{eqnarray}
C(N)&=&\int_1^NC_l(N)dl/N \nonumber\\
&\propto& \frac{N^{-2\beta-1}}{B^2(2\beta-1)^2}
\left\{N^{4\beta-1}[\frac{1}{2\beta}+\frac{2\beta-1}{4\beta-1}\ln N -\frac{2\beta-1}{(4\beta-1)^2}-\frac{2}{4\beta-1}]\right.\nonumber\\
 &&+N^{2\beta}(2\beta-3+\frac{3}{2\beta})-N^{2\beta-1}(2\beta-3+(2\beta-1)\ln
 N+\frac{1}{2\beta})\nonumber\\
&& \label{eq: avc}
\left.+\frac{2\beta-1}{(4\beta-1)^2}+\frac{2}{4\beta-1}-\frac{3}{2\beta}\right\}.
\end{eqnarray}
The above equation can be further simplified by keeping only the term with
the highest order of $N$, i.e., we have
%(\ref{eq: avc}) is a $\frac{0}{0}$ undefined type. Applying
%L'h\^{o}pital's rule twice to (\ref{eq: avc}), we obtain
\begin{equation}
%\lim_{\beta\rightarrow 1/2}
C(N)\propto N^{2\beta -2} \ln N.
\end{equation}
Obviously, we have
\begin{eqnarray}
\lim_{N\rightarrow \infty}C(N) \propto \lim_{N\rightarrow \infty}
N^{2\beta-2}\ln N =0. \label{eq: lim}
\end{eqnarray}

\begin{remark}
%From (\ref{eq: avc}), we observe that the clustering coefficient of our model is approximately $N^{2\beta-2}\ln N$.
Although the new model combines randomness and adaptability, the clustering
coefficient is still quite small for a relatively large $N$.
This shows that networks constructed with the two intrinsic rules
do not exhibit the small world property, although it does exists in
many real world scale-free networks.
What are the reasons for this inconsistency? What is missing in our construction process?
Our conjecture is that the third intrinsic evolution rule, i.e. {\bf hereditary},
has been ignored in the model. %inheritance
\end{remark}

\section{Average Path Length}
\ \ \ \ We now examine the relationship between the average short
path length $L$ and the total number of nodes $N$ in two
experiments. For each experiment, we set $m_0=10$, test the range
of $N$ from $10^3$ to $10^4$, and take the average from $100$
simulation runs. We then fit the data from the experiments by
linear regression.

Firstly, we examine the impact of $m$ and $p$ on $L$ by comparing
the following $5$ cases, with a fix $c=0$:

Case I: $m=8$, $p=0$;

Case II: $m=6$, $p=0$;

Case III: $m=4$, $p=0$;

Case IV: $m=4$, $p=0.4$;

Case V: $m=4$, $p=0.8$.

\begin{figure}[htbp]
\center{\psfig {file=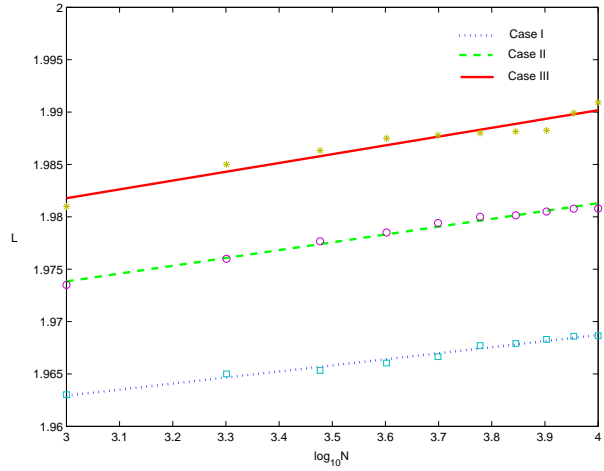, width=8cm}} \caption{ Impact of $m$
on $L$} \label{fig0}
\end{figure}

\begin{figure}[htbp]
\center{\psfig {file=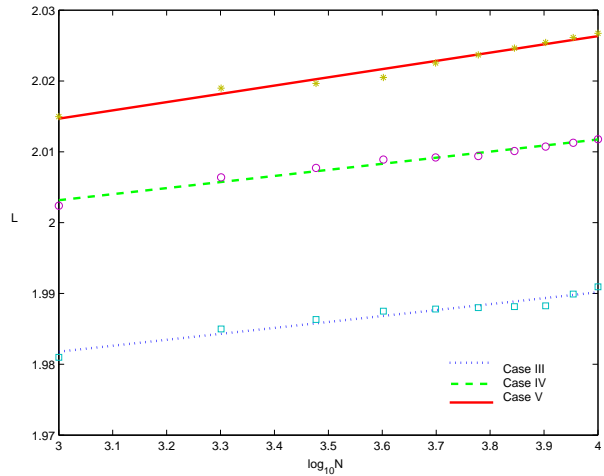, width=8cm}} \caption{ Impact of $p$
on $L$} \label{fig1}
\end{figure}

Comparing the first three cases, Figure \ref{fig0} shows that $L$
is decreasing in $m$. This is because the connectivity degree
increases with more newly added links. Figure \ref{fig1}
demonstrates how $L$ changes with $p$ in the last three
cases. We find that $L$ is increasing in $p$, which indicates that
the randomness results in the long $L$.

Secondly, we study the relationship between $L$ and $N$ under
different values of the degree exponent $\gamma$.

Case IV: $m=4$, $p=0.4$, $\gamma=3.167$;

Case VI: $m=5$, $c=1$ and $p=0.4$, $\gamma=2.667$.

\begin{figure}[htbp]
\center{\psfig {file=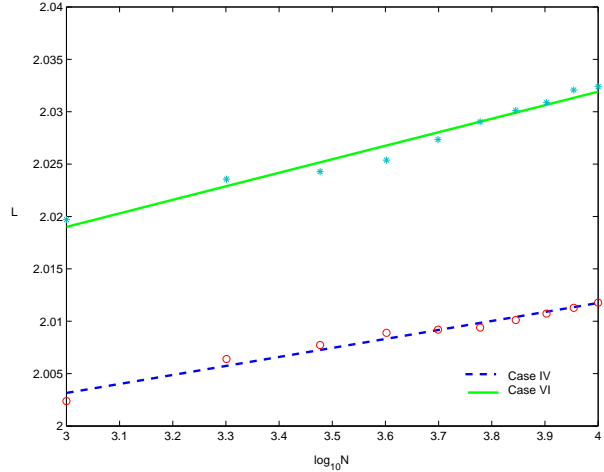, width=8cm}} \caption{ Impact of
$\gamma$ on $L$} \label{fig2}
\end{figure}

In Figure \ref{fig2}, we observe that $L$ in Case IV is a bit
shorter than that in Case VI. The phenomenon shows that although a
smaller $\gamma$ yields a larger probability that a node has more
links, it is not the only factor that determines $L$.
The length of $L$ also depends on the network construction mechanism.

Finally, noting that a log scale of the system size $N$ is used,
we can see in all three figures a logarithmic growth of $L$ with
respect to $N$.

\newpage

\section{Conclusions and Discussions}

\ \ \ \  There are two main contributions in this paper.
First, by successfully integrating randomness and adaptability,
we introduce a simple yet very flexible model for scale networks.
While, as demonstrated in the previous sections,
a number of the existing models are, in some way, special cases of our model,
we are still able to derive an explicit expression for the network
degree distribution.
%analytically calculate degree distributions of complex networks
%with preferential and random links, anti-preferential and random
%deletion. We show that the degree distribution follows a
%power law and the range of $\gamma$ lies in $(2,\infty)$.
%Thus, it self-organizes into a stationary scale-free network.
Our second contribution is the analytical expressions that we
obtain for the clustering coefficient for a large class of scale-free networks.
Apparently, there are not many successes in the literature for
cluster coefficients due to analytical difficulties.
Thus, the method we use in section 4 should be useful for others in the future.

Our discussion of cluster coefficients leads to an important observation,
i.e., Remark 3 in section 4.
Without hereditary, the important small world phenomenon
displayed in real networks cannot be captured in our model
as well as in many existing models.
This shows much remain to be done in our quest to understand complex networks better.

Some attempts have been made in including hereditary.
Ravasz and Barab$\acute{a}$si \cite{Ravasz02} build up a model of hierarchical organization with
deterministic copy of a module. Dorogovtsev et al.
\cite{Dorogovtsev01} model scale-free networks by a deterministic
pseudofractal graph. Although the authors show that the clustering
coefficient of a node follows a power law with respect to the
degree of the node, randomness and adaptability are absent.
Sol\'{e} et al. \cite{Sole02} investigate proteome growth model
with random node duplications, old removal edges, and newly added edges.
Empirically, they find by simulation that their model can
explain the macroscopic features exhibited by the proteome.
Klemm and Egu$\acute{\imath}$luz \cite{Klemm02}'s model combines the motif
copy and the BA model using a probability $\mu$. For $\mu=0$,
their model has the characters of small-world networks. For
$\mu=1$, their model is equivalent to the BA model.
But, due to the analytical difficulty for $0<\mu<1$, the performance
of the model can not be well studied. Holme and Kim \cite{Holme02}
integrate preferential attachment with triad information to
construct a scale-free network. By simulation, they show that
their model can generate small-world characters.

\end{document}